\documentclass[12pt]{article}

\def\lromn#1{\uppercase\expandafter{\romannumeral#1}}

\def\blist{\begin{list}{\setlength{\rightmargin}{\leftmargin}}}
\def\elist{\end{list}}

\def\bra#1{\langle #1 |}
\def\ket#1{| #1 \rangle}
\usepackage{graphicx}

\begin{document}

\begin{flushright}
 \today \\
 STUPP-04-177 \\
 ICRR-Report-507-2004-5
\end{flushright}

\begin{center}
\begin{large}

\textbf{
 Nonexponential decay of an unstable quantum system: \\
 Small-$Q$-value s-wave decay
 }

\end{large}

\vspace{2cm}

\begin{large}
 Toshifumi Jittoh$^a$, Shigeki Matsumoto$^b$, Joe Sato$^a$, Yoshio Sato$^a$ and Koujin Takeda$^c$ 

\vspace{0.5cm}

$^a$Department of Physics, Saitama University, \\
Saitama 338-8570, Japan \\
$^b$Institute for Cosmic Ray Research, University of Tokyo, \\
Chiba 277-8582, Japan \\
$^c$Department of Physics, Tokyo Institute of Technology, \\
Tokyo 152-8551, Japan
\end{large}

\vspace{3cm}

{\bf ABSTRACT}
\end{center}

We study the decay process of an unstable quantum system, especially
the deviation from the exponential decay law. We show that the
exponential period no longer exists in the case of the s-wave decay
with small $Q$ value, where the $Q$ value is the difference between the energy 
of the initially prepared state and the minimum energy of the continuous 
eigenstates in the system.  We also derive the quantitative condition that 
this kind of decay process takes place and discuss what kind 
of system is suitable to observe the decay.

\newpage

\lromn 1 \hspace{0.2cm} {\bf INTRODUCTION}
\vspace{0.5cm} 

Since the period of the classic works by Dirac \cite{Dirac} and
 Weisskopf and Wigner \cite{WWG},
it has been a problem how to describe the decay process of an unstable state
following the principles of quantum mechanics. As is well known,
the survival probability of the initial state $P(t)$, which concerns
the decay of the quantum state,
is frequently described by the exponential decay law
$P(t) = e^{-\Gamma t}$. 
However, it is also known
that the decay process does not obey the exponential law 
precisely,
 so it has always been a question of how
the deviation from the exponential decay law
occurs, especially at the late and early times 
of a decay process \cite{non-exp decay law}.

Theorists are always motivated to work on this old problem
when high-resolution experiments, which are
accomplished by a new technology, are performed
to detect the deviation of the decay law 
from the exponential \cite{test of e-decay 1} - \cite{test of e-decay 3}.
In addition, recent several
experiments have reported the measurement-induced suppression in quantum
systems at the early stage of decay,
which may be a result of the {\it Quantum Zeno
Effect} (QZE) \cite{QZE}.

As mentioned above, the deviation from the exponential law at late
and early times is often discussed \cite{deviation,SNE}.
 At very late times, the survival probability $P(t)$ must
decrease more slowly than the exponential and exhibits the
inverse power law of time $P(t) \sim t^{-\alpha}$, where $\alpha$ is
positive and depends on the property of the unstable system.
 At early times, 
 the survival probability decreases
 following a Gaussian law (the
square of time $t^2$), which appears inevitably in a quantum process
(and causes the QZE).
 Thus the decay of the unstable state proceeds through three
 stages in general. The initial stage is characterized by a
Gaussian law, the intermediate stage by an exponential law, and the
final stage by an inverse power law.

In this paper we focus on a different mechanism of deviation from
the exponential law. Such a decay process occurs in the case of 
small-$Q$-value s-wave decay (SQS decay).
Here the $Q$ value is defined by the difference between the energy 
of the initially prepared state (denoted by $E_0$) 
and the minimum energy of the continuous 
eigenstates (denoted by $E_{th}$) in the system. 
The small-$Q$-value decay has been discussed in
some papers \cite{SNE,smallQ}.
The point we would like to emphasize here is as follows: in the
case of the SQS decay, we can observe not only the
enhancement of the QZE, but also {\it no exponential period}. 
This means that
the deviation from the exponential law can be observed easily if
the SQS decay system is prepared. We also derive
the quantitative condition that such a decay takes place.

This paper is organized as follows. In Sec.\lromn 2, we show an example
of the SQS decay by using the one-dimensional
tunneling system with a box-type potential. The general description for an
unstable system is formulated in Sec.\lromn 3. Using this formalism,
we derive the quantitative condition that the SQS
decay occurs in Sec.\lromn 4. In Sec.\lromn 5, we summarize our results
and discuss what kind of system exhibits such a decay process.


\vspace{1cm}
\begin{center}
\lromn 2 \hspace{0.2cm}
{\bf AN EXAMPLE OF SQS DECAY IN TUNNELING PHENOMENA}
\end{center}
\vspace{0.5cm}

Before going into the general discussion, we show the example 
of SQS decay in the tunneling phenomena.
In this section we discuss the one-dimensional tunneling problem
 because only the radial part of the wave function is relevant
 to the s-wave tunneling even in a three-dimensional system.

Let us consider the decay process through the one-dimensional box-type potential depicted in Fig.1. Parameters for characterizing the system are also shown in the figure.
Here we assume $U_0$ is not so large that there is no bound state.
The goal here is to calculate the survival probability of the prepared state and show that the SQS decay is realized in this system. 

\begin{figure}
\begin{center}
  \includegraphics[width=7.5cm]{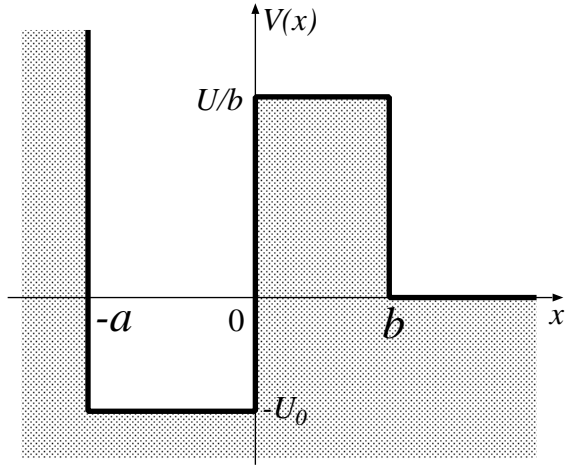}
\caption{\small 
The shape of the potential.
}
\end{center}
\end{figure}

The survival probability $P(t)$ is defined by the nondecay amplitude $a(t)$ as $P(t) = |a(t)|^2$. $a(t)$ is given by
\begin{eqnarray}
 a(t)
 \equiv
 \bra{0}e^{iHt}\ket{0}~,
\end{eqnarray}
where $\ket{0}$ is the initially prepared state and $H$ is the Hamiltonian
of the system. Please note that we use the units $\hbar=1$ in this paper. The nondecay amplitude can be expanded over the energy eigenstates $\ket{E}$ as
\begin{eqnarray}
 \label{survival}
 a(t)
 =
 \int_{0}^{\infty} dE~\rho(E)~e^{-iEt}~,
 \qquad
 \rho(E)
 =
 |\bra{E} 0 \rangle |^2~.
\end{eqnarray}
The function $\rho(E)$ is called the spectral function, in which all information about the decay process is included.
The energy eigenstate $\ket{E}$ can be obtained analytically in
 this system as detailed in Appendix A.
 Here we take the initial state $\ket{0}$ as the ground state
 in the well for the infinitely height barrier,
\begin{eqnarray}
\label{initial}
 \psi_i(x)
 \equiv
 \bra{x}0\rangle
 =
 \sqrt{\frac{2}{a}}\sin{\left(\frac{\pi x}{a}\right)}
 \theta(x + a)\theta(-x)~.
\end{eqnarray}
The energy expectation value of this state is $E_0 \equiv \pi^2/(2ma^2)-U_0$ 
and the $Q$ value is given by $Q \equiv E_0-E_{th}=\pi^2/(2ma^2)-U_0$ 
because the spectrum of the
 continuum energy eigenstate starts from the zero energy. Using this
 initial wave function, the spectral function of the system is obtained
 analytically after some calculations,
 and given by
\begin{equation}
\label{spectral}
 \rho(E) =
 \frac{1}{2ma^2\alpha(E)}\frac{2\pi q\sin^2r}{(r^2 - \pi^2)^2}~,
\end{equation}
where
\begin{equation}
 \alpha(E) = 
 q^2 + G_0\cos^2r - \frac{G}{u}\sin^2r
 + \frac{G}{u}
 \left(
  \frac{r}{s}\cos r\cdot\sinh(su) + \sin r\cdot\cosh(su) \right)^2\!\!\! , 
 \label{small} 
\end{equation}
for the case that the energy 
is smaller than the potential barrier $E \leq U/b$, while 
\begin{equation}
 \alpha(E) =
 q^2 + G_0\cos^2r - \frac{G}{u}\sin^2r
 + \frac{G}{u}
 \left(
  \frac{r}{\tilde{s}}\cos r\cdot\sin(\tilde{s}u)
  + \sin r\cdot\cos(\tilde{s}u)
 \right)^2. 
 \label{big}
\end{equation}
for the other case $E \geq U/b$.
Here we use the dimensionless quantities to write down the spectral function 
$q = \sqrt{2mE}a$, $r = \sqrt{2m(E + U_0)}a$, $s = \sqrt{2m(U/b - E)}a$, 
and $\tilde{s} = \sqrt{2m(E - U/b)}a$. 
Furthermore, the variables that characterize the potential 
are also given by the dimensionless ones $G = 2maU$, $G_0 = 2ma^2U_0$, 
and $u = b/a$. 

For investigating the SQS decay, we define the decay rate $\Gamma(t)$ by
\begin{eqnarray}
 \Gamma(t) = -\frac{d}{dt} \ln P(t)~.
\end{eqnarray}
This quantity is more convenient rather than the survival probability 
itself because this rate is constant while the decay process 
is governed by the exponential law. We performed the integral
in Eq.(\ref{survival}) numerically using the spectral function
 Eq.(\ref{spectral}), and calculated the decay rate 
$\Gamma(t)$. The results are shown in Fig.2. We studied 
two cases of $G_0$, that is, $G_0 = 0$ and $8.957 335$ which correspond
to the cases $Q= 8.973 65$ and $6.554 45\times 10^{-4}$ [the units of 
$Q$ are $(2ma^2)^{-1}$], with fixed 
$G$ and $U$.\footnote{
Rigorously speaking, there is a quantum correction to the
energy level and $Q$ should be given by eq.(26), where
$Q$ is indeed defined in this sense.
However, this correction is assumed to be small and
$Q$ is set to be $E_0-E_{th}$ in other lines or equations.
}
 As you see, the exponential decay law is observed 
in the case that the $Q$ value is not very small. 
On the other hand, if the $Q$ value is small enough
the exponential period no longer exists even
at the time $t \sim 500 (2ma^2)$ when $P(t)$ decreases
 to the order of $e^{-1}$. This is nothing but an example 
of the SQS decay.

\begin{figure}
\begin{center}
  \includegraphics[width=15cm]{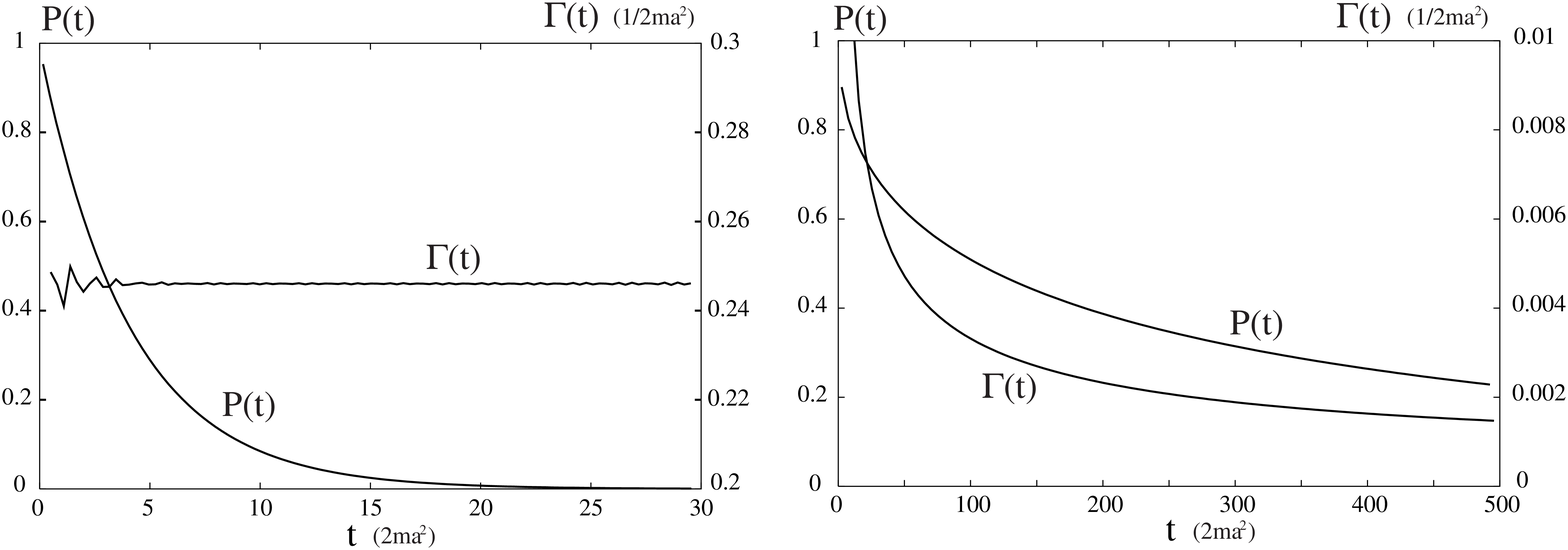}
\caption{
\small Examples of decay. Here we set 
 $G=20, G_0=0$ (or $Q=8.973 65(2ma^2)^{-1} $),
 $u=10^{-4}$ on the left and
 $G=20, G_0=8.957 335$ (or $Q=6.554 45 \times 10^{-4}(2ma^2)^{-1}$),
 $u=10^{-4}$ on the right.  
 Each quantity in the figure is averaged over a short time interval.
 The example on the right corresponds to the SQS decay.}
\end{center}
\end{figure}

In the following sections we investigate the SQS decay by a general
description and what kind of situation is necessary. We also derive the 
quantitative condition for the SQS decay to take place.


\vspace{1cm}
 \lromn 3 \hspace{0.2cm}
 {\bf GENERAL DESCRIPTION FOR UNSTABLE STATE}
\vspace{0.5cm}

In this section we explain the general formalism of unstable state decay.
Using this formalism the quantitative condition for the SQS decay is
 derived in the next section. 

For calculating the nondecay amplitude we use an 
exact integro-differential equation by using the technique in 
Refs. \cite{Peres,SNE}. 
We introduce the projector onto the initial unstable state, 
${\cal P} \equiv \ket{0}\bra{0}$, and decompose the Hamiltonian $H$ as
\begin{eqnarray}
 H &=& H_0 + V~,
 \nonumber \\
 H_0 &\equiv& {\cal P}H{\cal P} + (1 - {\cal P})H(1 - {\cal P})~, 
 \qquad
 V \equiv H - H_0~.
\end{eqnarray}
Define the energy eigenstates $\ket{a}$ of $H_0$;
$H_0\ket{a} = E_a\ket{a}$.
It is then easy to confirm that the
decay interaction $V$ operates only between
the initial state $\ket{0}$ and its orthogonal complement $\ket{n}$;
$V_{0n} = V_{n0}^* \neq 0$, $V_{mn} = V_{00} = 0$.
Here we use the intermediate roman letters such as $m$, $n$
to denote eigenstates projected by $1 - {\cal P}$.
The interaction $V$ depends on the initially prepared state.
Although this formalism looks odd, we can execute an exact analysis
like a time development of the unstable state using this tool.

We work in the interaction picture and expand the state at a finite time $t$,
 using the basis of the eigenstate of $H_0$; 
$\ket{\psi}_I = e^{iH_0t}\ket{\psi} = \sum_a c_a(t) \ket{a}$.
We thus write the time evolution equation for the coefficient $c_{a}(t)$:
\begin{eqnarray}
 i\dot{c}_0
 =
 \sum_n~V_{0n}~e^{-i(E_n - E_0)t}~c_n~,
 \qquad
 i\dot{c}_m
 =
 V_{m0}~e^{i(E_m - E_0)t}~c_{0}~.
\end{eqnarray}
Here $E_0$ is the energy of the initial unstable state;
$E_0 = \bra{0}H\ket{0}$.
The nondecay amplitude is related to this coefficient by
$\bra{0}e^{-iHt}\ket{0} = e^{-iE_0t}c_0(t)$. From the above
equations, a closed form of the equation for the nondecay amplitude
$c_0(t) \equiv a(t)$ then follows:
\begin{eqnarray}
 \dot{a}(t)
 &=&
 -\int_0^t~dt~\beta (t - t')~a(t')~,
 \label{exact int-diff eq} 
 \\
 \beta (t - t')
 &=&
 \bra{0} V_I(t)V_I(t')\ket{0}
 =
 \int_{E_{th}}^\infty~dE~
 \sigma (E)~e^{-i(E - E_0)(t - t')}~,
 \\
 \sigma (E)
 &=&
 \sum_m \delta (E - E_m) |V_{0m}|^2~.
\end{eqnarray}
Here $V_I(t) = e^{iH_0t}~V~e^{-iH_0t}$
is the decay interaction written in the interaction picture and
the function $\sigma (E)$ characterizes the interaction between the
unstable state $\ket{0}$ and the other states $\ket{m}$.
The initial condition $c_m(0) = 0$ is used to derive the equation for
$a(t)$, and $E_{th}$ is the threshold for the state $\ket{m}$.

The standard technique to solve this type of integro-differential equation
(\ref{exact int-diff eq}) is the one that utilizes the Laplace transform,
and we finally obtain the nondecay amplitude in the form
\begin{eqnarray}
 a(t)
 &=&
 \int_{-\infty}^\infty \frac{dE}{2\pi i}
 F(E + i0^+)~e^{-i(E - E_0)t}~,
 \label{amp by analytic fn}
 \\
 F(z)
 &\equiv&
 \frac{1}{- z + E_0 - G(z)}~,
 \qquad
 G(z)
 =
 \int_{E_{th}}^\infty dE
 \frac{\sigma (E)}{E - z}~.
\end{eqnarray}
The initial condition $a(0) = 1$ is imposed in this derivation.

The analytic property of the function $F(z)$
is evident; this function is analytic
except on the branch
cut which runs from the threshold value $E_{th}$ to 
positive infinity on the real axis (Fig.3). As is well known,
if the Riemann surface is considered by analytic continuation through the
branch cut
(and regarding the original complex $z$ plane as the first Riemann sheet),
 there is a pole on the {\it second} Riemann sheet
 near and below the real axis
if the decay interaction is weak enough.
The pole location $z_p$ is determined by
\begin{equation}
 z_p - E_0 + G_{\lromn 2}(z_p)
 =
 z_p - E_0
 +
 \int_{E_{th}}^\infty dE
 \frac{\sigma (E)}{E - z_p}
 +
 2\pi i\sigma(z_p)
 =
 0~,
 \label{pole location}
\end{equation}
where the analytic function $G(z)$, and hence $F(z)$, 
is extended into the second sheet by
$G_{\lromn 2}(E - i0^{+}) = G_{\lromn 1}(E + i0^{+})$
through the branch cut.
The real function $\sigma(E)$, which was originally defined
 for real $E>E_{th}$,
 is also extended to the function $\sigma(z)$ defined on the complex plane
 by analytic continuation.
In addition to this pole there may be some singularities on the 
second Riemann sheet, but 
we ignore the effects of such singularities in the following discussion.
This approximation is valid as discussed in Ref. \cite{SNE} 
because these singularities do not affect the decay phenomena except 
for the early stage. 

Using the discontinuity of the analytic function $F(z)$ across the
 branch cut on the first Riemann sheet,
\begin{equation}
 F(E + i0^+) - F(E - i0^+)
 =
 2\pi i\sigma (E)~|F(E + i0^+)|^2~,
\end{equation}
which is called the "elastic" unitarity relation, we can deform
the contour of integration on the real axis in Eq.(\ref{amp by analytic fn})
into the sum of two contours,
one around the pole as shown by $C_p$ 
and the other along $C_{th}$ in Fig.3,
\begin{equation}
 a(t)
 =
 \left(
  \int_{C_p} + \int_{C_{th}}
 \right)
 \frac{dz}{2\pi i} F(z)~e^{-i(z - E_0)t}~.
 \label{amp by 2-cont}
\end{equation}

\begin{figure}
\begin{center}
  \includegraphics[width=7.5cm]{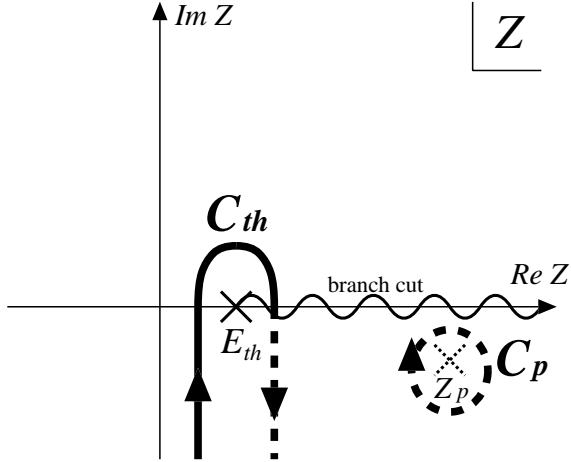}
\caption{\small 
The analytic structure of the complex $z$ plane 
and the contours of the integrals. The
contours and the pole shown by the broken line are on the second Riemann sheet. 
}
\end{center}
\end{figure}

We consider the case that the unstable state initially prepared
is a metastable state, which means the decay interaction is weak 
in comparison with the typical energy of the system, $E_{sys}$, induced
from the oscillation in the well: $\sigma \ll E_{sys} \sim E_0$.
Then the pole location on the second Riemann sheet can be obtained
approximately as
\begin{eqnarray}
\label{app zp}
 z_p 
 \simeq
 E_0 - \Pi(E_0) - i\pi \sigma(E_0)~,
 \qquad
 \Pi(E)
 =
 {\rm p.v.}\int_{E_{th}}^\infty dE'
 \frac{\sigma (E')}{E' - E}~,
\end{eqnarray}
where p.v. means the principal value of the integration.
The $C_p$ integration in Eq.(\ref{amp by 2-cont})
can be performed without difficulty by the residue theorem, 
and with the aid of the approximation above, this becomes
\begin{eqnarray}
 \int_{C_p}\frac{dz}{2\pi i} F(z)~e^{-i(z - E_0)t}
 \simeq
 e^{i\Pi(E_0)t}e^{-\pi \sigma(E_0) t}~.
 \label{Cp-integration}
\end{eqnarray}
This integration gives an ${\cal O}(1)$ contribution
to the nondecay amplitude $a(t)$.
On the other hand, the integration along $C_{th}$
is of ${\cal O}(\sigma/E_0)$,
 which gives only a small contribution.
This is because the integration can be approximated by
\begin{eqnarray}
 \int_{C_{th}} \frac{dz}{2\pi i} F(z)~e^{-i(z - E_0)t}
 &=&
 -ie^{i(E_0 - E_{th})t}\int^\infty_0 dy~
 \sigma(E_{th} - iy)|F(E_{th} - iy)|^2~e^{-yt}~,
 \nonumber \\
 &\simeq&
 -ie^{i(E_0 - E_{th})t}|F(E_{th})|^2
 \int^\infty_0 dy~\sigma(E_{th} - iy)~e^{-yt}~,
 \label{Cth-integration}
\end{eqnarray}
for sufficiently large $t$, and the factor $|F(E_{th})|^2$
usually takes the value $(E_0-E_{th})^{-2}$.

The dominance of the $C_p$ integration
in Eq.(\ref{amp by 2-cont}) leads to the exponential decay law 
\begin{eqnarray}
 P(t)
 =
 |a(t)|^2 \simeq \exp(-\Gamma_p t)~,
 \qquad
 \Gamma_p
 =
 2\pi\sigma(E_0)~,
\end{eqnarray}
and this coincides with the familiar golden rule of perturbation
theory. From this investigation,
we know that the perturbative calculation gives
satisfactory results in many cases.
We would, however, like to elucidate the time evolution in finer detail,
and study the conditions for breaking the exponential decay law, especially 
the SQS decay, in the next section.


\vspace{1cm}
 \lromn 4 \hspace{0.2cm}
 {\bf THE CONDITION FOR THE SQS DECAY LAW}
\vspace{0.5cm}

We now derive the quantitative condition of SQS decay. 
We also discuss the situations in general that 
exhibit deviations from the exponential decay law. 

The $C_p$ integration around the pole $z_p$ in Eq.(\ref{amp by 2-cont})
always yields exponential time dependence, 
so nonexponential decay is realized
when the $C_{th}$ integration contributes to the nondecay amplitude
$a(t)$ by the same order as the $C_p$ integration.

From the evaluation of the contour integrations in Eq.(\ref{amp by 2-cont}),
we can classify the nonexponential decays that satisfy the
condition mentioned above into three cases.
In the following we describe them in detail.

The first one concerns the short time behavior 
and is known as the QZE.
At early times ($t \leq E_{sys}^{-1}$), the approximation used in
Eq.(\ref{Cth-integration})
is no longer valid and the high-frequency component of $\sigma (E)$
 becomes important.
From the definition of the survival probability, 
we naively expect that the short time behavior exhibits a deviation
from the exponential law, which is in the form of
\begin{equation}
 |\bra{0}e^{-iHt}\ket{0}|^2
 \simeq
 1 - t^{2}\,
 \left(
  \bra{0}H^2\ket{0} - \bra{0}H\ket{0}^2
 \right)~.
 \label{short-time}
\end{equation}
Thus quantum mechanics appears to predict
a quadratic form of deviation in the $t\rightarrow 0$ limit.

The second one relates to the long time behavior.
At late times $t\gg 1/\Gamma_p$ the $C_p$ integration
is exponentially suppressed, while the $C_{th}$ integration is not
strongly suppressed because the behavior of $\sigma (E)$ near the
threshold is expressed by $\sigma (E) =  c(E - E_{th})^\alpha$,
 which leads to power law behavior of the $C_{th}$ integration,
\begin{eqnarray}
 \int_{C_{th}} \frac{dz}{2\pi i} F(z)~e^{-i(z - E_0)t}
 &\simeq&
 -ie^{i(E_0 - E_{th})t}|F(E_{th})|^2
 \frac{c~\Gamma(\alpha + 1)~e^{-i\pi\alpha/2}}{t^{\alpha + 1}} ~.
 \label{Large-time}
\end{eqnarray}
Here $\Gamma(z)$ is Euler's gamma function.
Therefore the contribution from the $C_{th}$ integration 
exceeds the one from the $C_p$ integration
at late times, and the decay law changes from exponential to
an inverse power law.

The small $Q~(= E_0 - E_{th})$ value case is the last one
on which we shall mainly focus in this paper,
that is, the SQS decay.
This case can be understood by investigation of the prefactor
$|F(E_{th})|^2$ in Eq.(\ref{Cth-integration}) in detail:
\begin{eqnarray}
 |F(E_{th})|^2
 =
 \left[Q - \Pi(E_{th})\right]^{-2}~.
\end{eqnarray}
Since the function $\Pi(E_{th})$ is of ${\cal O}(\sigma)$, 
this factor gives $Q^{-2}$ when the $Q$ value is not
very small. However, if the $Q$ value is of the same order as $\sigma$,
the factor $|F(E_{th})|^2$ becomes large and
the contribution from the $C_{th}$ integration becomes 
comparable with the one from the $C_p$ integration.
In this case decay that does not include
the exponential period at all can be realized, and this
situation never occurs in other cases described above.

We now derive the quantitative condition that the 
SQS decay takes place.
By virtue of the analytic property of $F(z)$, 
we can express the amplitude $a(t)$ in the convenient form
\begin{eqnarray}
 a(t)
 =
 \int_{E_{th}}^\infty dE \rho(E)e^{-i(E - E_0)t}~,
 \qquad
 \rho(E)
 = 
 \frac{\sigma (E)}
 {(E - E_0 + \Pi(E))^2 + (\pi \sigma (E))^2}~.
 \label{amp with rho} 
\end{eqnarray}
The function $\rho(E)$ is the spectral function as mentioned in Sec.\lromn 2.
The schematic shape of $\rho(E)$ is shown in Fig.4. This function takes a 
real positive value when $E$ is larger than
the threshold $E_{th}$.
$\rho(E)$ drops quickly in the limit $E\rightarrow\infty$
because this function must satisfy the normalization condition
$\int^\infty_{E_{th}}dE \rho(E) = 1$, which is equivalent to  
the initial condition $a(0) = 1$. In addition, 
$\rho(E)$ has a peak around $E \sim E_0$.
(More precisely, 
the exact location and the width of the peak
are determined by the real and the imaginary parts of $z_p$
in Eq.(\ref{pole location}).)
Therefore the dominance of the $C_p$ integration
is equivalent to the Breit-Wigner form of $\rho(E)$
 or the limit $E_{th} \rightarrow -\infty$, as
you can find from the shape of $\rho(E)$.

\begin{figure}
\begin{center}
  \includegraphics[width=7.5cm]{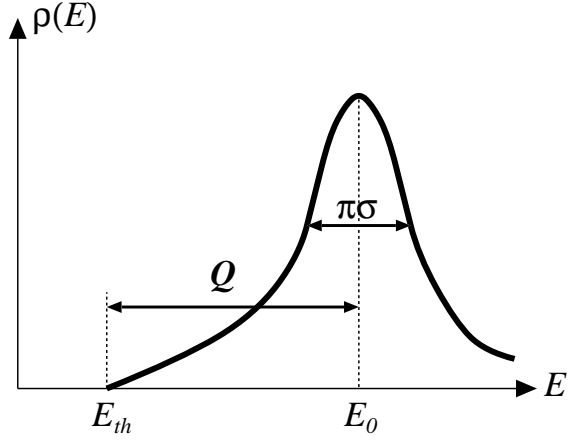}
\end{center}
\caption{
\small
The typical shape of the spectral function $\rho(E)$.
 The SQS decay is realized when the $Q$ value is smaller than
 the width of the peak.}
\end{figure}

The absence of the exponential decay law due to the small $Q$ value 
then occurs when $Q$ is smaller than the width of the peak
because the Breit-Wigner shape of $\rho(E)$ does not
hold in this situation.
Thus the condition for the SQS decay is given by
\begin{eqnarray}
 \frac{|\Im z_p|}{Q} \geq 1~,
 \qquad
 Q = \Re z_p - E_{th}~.
 \label{criterion}
\end{eqnarray}
This is the final result of this section, the quantitative condition for
the SQS decay to take place.
In Ref. \cite{SNE}, the authors derived the opposite condition to
Eq.(\ref{criterion}) from the viewpoint that nearly exponential 
decay takes place. 
The condition here is a necessary and sufficient condition, so our
condition is consistent with theirs.

To investigate the condition Eq.(\ref{criterion}) for a given system,
this inequality is not very useful because
the calculation for the exact location of 
$z_p$ from Eq.(\ref{pole location}) is a tedious one.
This situation is somewhat improved
if we use the approximation Eq.(\ref{app zp}); then the 
condition turns to
\begin{equation}
 \frac{\Gamma_p}{E_0-E_{th}} \geq 2~. 
\end{equation}
This is more convenient for practical use.

As an example, we show the contour plot of $\Im z_p/Q$ 
for the model discussed in Sec.\lromn 2, which is depicted in Fig.5. 
The contour plot is drawn on the $(G,Q)$ plane with fixed $u$. 
The crosses in the figure correspond to the values used in Fig.2, and 
in the shaded region the parameters satisfy the 
condition that the SQS decay takes place.
\begin{figure}
\begin{center}
  \includegraphics[width=7.5cm]{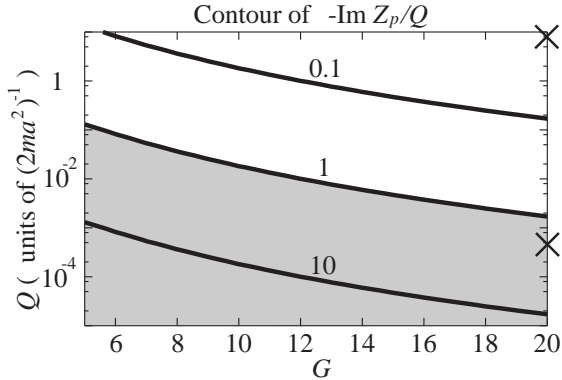}
\caption{
\small
The value of Eq.(\ref{criterion}) 
is depicted as a contour map in the $(G,Q)$ plane.}
\end{center}
\end{figure}


\vspace{0.5cm}
\lromn 5 \hspace{0.2cm}
{\bf DISCUSSION AND SUMMARY}
\vspace{0.5cm} 

We now come to the stage of discussing what kind of system
 is necessary for the SQS decay.
The important point we must pay attention to for discussing
the condition Eq.(\ref{criterion}) is the
$Q$ dependence of $\Im z_p$, the imaginary part of the pole location.
 The $Q$ dependence is determined by the function $\sigma(E_0)$
as shown in Eq.(\ref{app zp}).
 When the $Q$ value is small enough, the dependence
is determined by the threshold behavior of $\sigma(E_0)$, which 
originates from the spectral function 
$\rho(E_0)$ as shown in Eq.(\ref{amp with rho}).
Therefore the threshold behavior of the spectral function, 
taken to be $\rho (E_0) =  c(E_0 - E_{th})^\alpha
= cQ^\alpha$, is the key quantity in this problem. Here the
 coefficient $c$ is a constant dependent on the system. The SQS decay is
expected when the power of $\rho(E_0)$ near the threshold, $\alpha$,
is smaller than 1.

As is well known, the threshold behavior of the spectral function is
determined by the quantum number of the orbital angular momentum
in a scattering process (e.g., a particle decay or 
a radioactive process), which we denote $l$ \cite{Landau}.
The threshold behavior is given by 
$\rho(E_0) = c(E_0 - E_{th})^{l + 1/2}$. 
Therefore {\it s-wave} ($l = 0$) decay is necessary 
for the SQS decay in these processes, and the SQS decay 
never occurs via higher-$l$ ($l \geq 1$) processes.

Let us move on to the decay process through tunneling. 
For the one-dimensional model discussed in Sec.\lromn 2,
the threshold behavior is given by
\begin{eqnarray}
 \rho(E_0)
 =
 \frac{Q^{1/2}}{4\pi ma^2\alpha(0)}~,
 \qquad
 Q = E_0 - E_{th}~,
\end{eqnarray}
which is the same as in the case of the s-wave decay.
This is the very reason that the system exhibits the
SQS decay when the $Q$ value is small enough.
You might think that such a threshold behavior is due to a
peculiarity of the potential. This is, however, not correct.  To check this, 
let us consider a system with the modified potential 
shown in Fig.6. The spectral function of this system can also be 
obtained analytically using Bessel functions. 
The threshold behavior again coincides with the case
of the s-wave decay. 
The form of the spectral function and the threshold behavior 
are given in Appendix B.

\begin{figure}
\begin{center}
  \includegraphics[width=7.5cm]{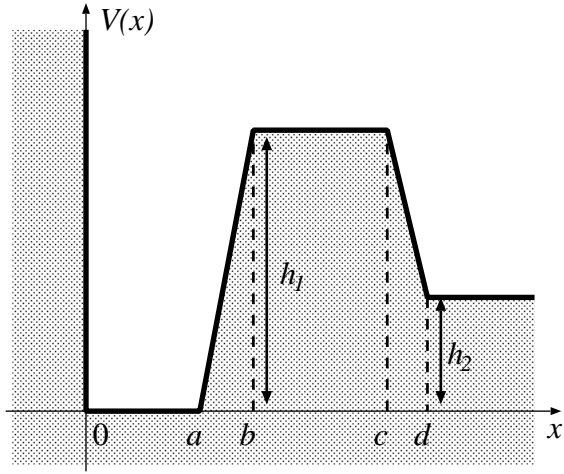}
\caption{
\small
The modified potential.}
\end{center}
\end{figure}

This result is naturally understood if we consider 
the model with spherical symmetry in three dimensions. 
Since the angular momentum is a conserved quantity in this case, 
the s-wave decay process can always be reduced to a
problem in a one-dimensional system. 
Therefore the threshold behaviors of all models in one dimension 
with nonsingular potential are the same as 
the one of s-wave decay.

We summarize the results of this paper.
The decay of the unstable state with small $Q$
value in the {\it s-wave} process (SQS decay)
exhibits the interesting feature that there is  
{\it no exponential period}.
If we can make practical use of this mechanism, 
we will easily observe the deviation from the exponential law.
As mentioned above, the SQS decay takes place in a
system described by an s-wave process or 
a one-dimensional system. However
it is difficult to prepare a setup of small $Q$ value
in experiments on particle decay 
or radioactive processes, 
because the $Q$ values in such cases are fixed by nature and
we cannot control them.
On the other hand, the tunneling
phenomenon may be hopeful to observe the SQS process because  
we may achieve sufficiently small $Q$ more easily.
Unfortunately, experiments to look for the SQS decay have not been carried out
 until now. It is important to discuss an actual physical system 
which realizes the SQS decay, 
but this issue is beyond the scope of this paper and remains 
as a future problem.
We believe that this kind of experiment is interesting 
to observe the nonexponential decay of an unstable quantum system.

\begin{center}
{\bf Acknowledgments}
\end{center}
The authors would like to kindly thank Professor S. Takagi 
for discussion.
J.S. was supported by the Grants-in-Aid for Scientific Research
 on Priority Areas No. 16038202 and No. 14740168.
 K.T. was supported by the 21st Century COE Program at Tokyo Institute of
 Technology ``Nanometer-Scale Quantum Physics''. 


\vspace{1cm}
\begin{center}
  {\bf APPENDIX A}
\end{center}
\vspace{0.5cm}

In this appendix, we calculate the spectral function 
$\rho(E) = |\bra{E}0\rangle|^2$ of the one-dimensional model
with a box-type potential that is used in Sec.\lromn 2.
The explicit form of the energy eigenstate $\ket{E}$ 
is necessary in this computation and it can be obtained analytically,
\begin{eqnarray}
 \phi_E(x)
 =
 \bra{x}E\rangle
 =
 \left\{
  \begin{array}{ll}
   A \sin r(x/a + 1)
   &
   $for$ \quad (-a \leq x \leq 0)~,
   \\
   B \sinh (sx/a) + C \cosh (sx/a)
   & 
   $for$ \quad (0 < x < b),
   \\
   D \sin (qx/a) + E \cos (qx/a)
   &
   $for$ \quad (b \leq x),
  \end{array}
 \right.
\end{eqnarray}
for $E < U/b$ and
\begin{eqnarray}
 \phi_E(x)
 =
 \bra{x}E\rangle
 =
 \left\{
  \begin{array}{ll}
   A \sin r(x/a + 1)
   &
   $for$ \quad (-a \leq x \leq 0)~,
   \\
   B \sin (\tilde{s}x/a) + C \cos (\tilde{s}x/a)
   & 
   $for$ \quad (0 < x < b),
   \\
   D \sin (qx/a) + E \cos (qx/a)
   &
   $for$ \quad (b \leq x),
  \end{array}
 \right.
\end{eqnarray}
for $E \geq U/b$.
All parameters such as $a$, $b$, $q$, $r$, $s$, and $\tilde{s}$ are defined in
Sec.\lromn 2. The coefficients $A$, $B$, $C$, $D$, and $E$ can be determined by the
junction condition at $x=-a$, $0$, $b$ and the normalization condition of the
eigenstates, $\bra{E}E'\rangle = \delta(E - E')$. For example, the coefficient
$A$ is given by
\begin{eqnarray}
 |A|^2
 =
 \frac{q}{2ma^3\pi\alpha(E)}~,
\end{eqnarray}
and others can be obtained similarly.
The function $\alpha(E)$ is defined in Eqs.(\ref{small}) and (\ref{big}).
The spectral function is defined by the overlap of the energy eigenstate 
$\ket{E}$ and the initially prepared state $\ket{0}$ given in
Eq.(\ref{initial}). Using the eigenstate obtained above,
the spectral function is computed as
\begin{eqnarray}
 \rho(E)
 =
 |\bra{E}0\rangle|^2
 =
 \left|
  \int dx \phi_E(x)^*\psi_i(x)
 \right|^2
 =
 \frac{1}{2ma^2\alpha(E)}\frac{2\pi q\sin^2r}{(r^2 - \pi^2)^2}~.
\end{eqnarray}


\vspace{1cm}
\begin{center}
  {\bf APPENDIX B}
\end{center}
\vspace{0.5cm}

In this appendix, we write down the explicit form of the spectral function 
of the model whose potential is depicted in Fig.6 and
which is used in the discussion of Sec.\lromn 5.
The spectral function of this system can also be obtained analytically
using Bessel functions, and the computation of the spectral function can
be performed in the same way as in Appendix A. 
After some calculations we obtain
\begin{equation}
 \rho(E)
 =
 \frac{1}{2ma^3k_3(|\epsilon_1|^2+|\epsilon_2|^2)}
 \frac{\pi \sin^2(k_1a)}{(k_1^2a^2-\pi^2)^2}~,
 \label{daikei}
\end{equation}
where $k_1 = \sqrt{2mE}$, $k_2 = \sqrt{2m(h_1 - E)}$, and
$k_3 = \sqrt{2m(E - h_2)}$. All parameters characterizing 
the potential such as
$a, \ldots, d$ and $h_1, h_2$ are defined in Fig.6. The variables
$\epsilon_1$ and $\epsilon_2$ are given by
\begin{eqnarray}
 \left(
  \begin{array}{c}
   \epsilon_1
   \\
   \epsilon_2
  \end{array} 
 \right)
 =
 F^{-1}(k_3)
 \cdot
 T_R(z_d)
 \cdot
 T^{-1}_R(z_c)
 \cdot
 Ph
 \cdot
 F^{-1}(-ik_2)
 \cdot
 T_L(z_b)
 \cdot
 T^{-1}_L(z_a)
 \left(
  \begin{array}{c}
   \delta_1
   \\
   \delta_2
  \end{array}
 \right)~.
 \label{epsilon12}
\end{eqnarray}
$L$ and $R$ appearing as the subscripts of the matrix $T$
represent the slopes of the barrier at the left and the right sides,
$L = 2ma^3h_1/(b - a)$, $R = 2ma^3(h_1 - h_2)/(d - c)$.
The matrices $F(k)$ and $T_t(z)$ are defined by
\begin{eqnarray}
 F(k)
 \equiv
 \left(
  \begin{array}{cc}
   1 & 1
   \\
   ik a & -ik a
  \end{array} 
 \right)~,
 \qquad
 T_t(z)
 \equiv
 \left( 
  \begin{array}{cc}
   B_{1/3}(z) & B_{-1/3}(z)
   \\
   t^{1/3}B'_{1/3}(z) & t^{1/3} B'_{-1/3}(z)
  \end{array} 
 \right)~.
\end{eqnarray}
The arguments of the matrix, $z_a, \ldots, z_d$ are
the rescaled positions of $a, \ldots, d$, which are defined by
$z_a = -L^{1/3}(b - a)E/(ah_1)$, 
$z_b = L^{1/3}(b - a)(h_1 - E)/(ah_1)$,
$z_c = R^{1/3}(d - c)(h_1 - E)/(a(h_1 - h_2))$,
$z_d = R^{1/3}(d - c)(h_2 - E)/(a(h_1 - h_2))$.
The function $B_{\pm 1/3}(z)$ is defined by
the Bessel function $J_{\pm 1/3}$ as
\begin{equation}
 B_{\pm 1/3}(z) \equiv \frac{\sqrt{-\pi z}}{3}
 J_{\pm 1/3} \left( \frac{2}{3}(-z)^{3/2}\right)~,
\end{equation}
and $B'(z)$ means the derivative with respect to $z$.
$\delta_1$ and $\delta_2$ at the right side in Eq.(\ref{epsilon12})
are the values of the wave function and its derivative at the origin,
which are given by $\delta_1 = \sin (k_1a)$,
$\delta_2 = k_1a \cos (k_1a)$. Finally, the matrix $Ph$ is given by
\begin{eqnarray}
 Ph
 \equiv
 \left(
  \begin{array}{cc}
   e^{k_2(c - b)} & e^{-k_2(c - b)}
   \\
   k_2a e^{k_2(c - b)} & k_2a e^{-k_2(c - b)}
  \end{array} 
 \right)~.
\end{eqnarray}

In the rest of the appendix we show that the threshold behavior 
of this system is same as that of the s-wave decay. 
We expand the spectral function with respect to $(E - h_2)$
in the vicinity of the threshold.
The components of the matrices
in Eq.(\ref{epsilon12}),  $T_R(z_d)$, 
$T^{-1}_R(z_c)$, $Ph$, $F^{-1}(-ik_2)$,
$T_L(z_b)$, and $T^{-1}_L(z_a)$, and the variables 
$\delta_1$,$\delta_2$ become 
constant at the leading order, and only the
matrix $F^{-1}(k_3)$ has $E$ dependence such that
\begin{eqnarray}
 F^{-1}(k_3)
 \simeq
 \frac{1}{2}
 \left(
  \begin{array}{cc}
   1 & +ik_3^{-1}
   \\
   1 & -ik_3^{-1}
  \end{array}
 \right)~.
\end{eqnarray}
Therefore $\epsilon_1$ and $\epsilon_2$ are proportional to 
$k_3^{-1}$. As a result, the threshold behavior of 
the spectral function
$\rho(E)$ of Eq.(\ref{daikei}) is of the form
$\rho(E) \propto k_3 = (E - h_2)^{1/2}$,
which is same as the result of s-wave decay.


\end{document}